\documentclass[pdflatex,sn-mathphys]{sn-jnl}

\usepackage{xcolor}
\usepackage{soul}

\jyear{2023}%

\theoremstyle{thmstyleone}%
\theoremstyle{thmstyletwo}%

\theoremstyle{thmstylethree}%

\begin{document}

\title[Surgical Guidance]{Three-Dimensional Sonification as a Surgical Guidance Tool}

\author*[1,2]{\fnm{Tim} \sur{Ziemer}}\email{tim.ziemer@uni-hamburg.de}

\affil*[1]{\orgdiv{Institute of Systematic Musicology}, \orgname{University of Hamburg}, \orgaddress{\street{Neue Rabenstr. 13}, \postcode{20354} \state{Hamburg}, \country{Germany}}}

\affil[2]{\orgdiv{Bremen Spatial Cognition Center}, \orgname{University of Bremen}, \orgaddress{\street{Enrique-Schmidt-Str. 5}, \postcode{28359 } \state{Bremen}, \country{Germany}}}


\abstract{Interactive Sonification is a well-known guidance method in navigation tasks. Researchers have repeatedly suggested the use of interactive sonification in neuronavigation and image-guided surgery. The hope is to reduce clinicians' cognitive load through a relief of the visual channel, while preserving the precision provided through image guidance. In this paper, we present a surgical use case, simulating a craniotomy preparation with a skull phantom. Through auditory, visual, and audiovisual guidance, non-clinicians successfully find targets on a skull that provides hardly any visual or haptic landmarks. The results show that interactive sonification enables novice users to navigate through three-dimensional space with a high precision. The precision along the depth axis is highest in the audiovisual guidance mode, but adding audio leads to higher durations and longer motion trajectories.}



\maketitle

\section{Introduction}
\label{intro}
Image guided surgery is an umbrella term for surgical interventions where planing and execution are supported by pre-operative patient images, like magnetic resonance imaging or computed tomography \cite{nguide,igs,igs2}. This is necessary for complicated interventions, for example, when a tumor lies close to a risk-structure, such as a large vessel, sensitive membrane, or important nerve, or when an ablation needle has to reach a tumor past impenetrable bones. To plan the intervention, the image is segmented into parts that belong together, like muscles, vessels, bones, nerves, tumor and surrounding healthy tissue. Segmented images are augmented, e.g., by giving structures individual colors and opacity levels. With the help of such an augmented, three-dimensional image, the surgical procedure is planned. For example, the locations of burr holes are planned such that a very specific part of the skull can be removed for a craniotomy. 
In many interventions, visual, computer assisted guidance tools are utilized: the position of the surgical tool is being tracked in relation to the patient anatomy. A computer screen displays a pseudo-3D image of the augmented patient anatomy, overlaid by the planned points or path, and a depiction of the surgical tool.

Auditory display has been suggested as an auditory navigation aid for computer assisted surgical procedures for decades \cite{vrsurg,whatwedo,vickers,ijis,davidnewsletter,implant}. It has been recognized that computer assisted navigation enables clinicians to carry out complicated interventions with a high precision, even in minimally invasive surgery, where there is little or no direct view on the lesion. One shortcoming of visual guidance is that visual attention is captured by the screen instead of the patient. This can cause an unergonomic posture, may potentially reduce the precision of the natural hand-eye coordination, and extracting three-dimensional navigation information from a two-dimensional computer screen causes a high cognitive load \cite{tina}. Interactive sonification has the potential to overcome these issues. 

Experiments under laboratory conditions \cite{igic,davidneedle,davidstandalone,surgrev2,davidjournal,Plazak2019,surgdistson} as well as case reports of clinical implementations \cite{surgerycoll,surgicalnavigationau,otologic} exist. An overview is provided in \cite{Black2017}. These studies have led to similar conclusions: interactive sonification is a successful guidance tool for image guided surgery. Auditory guidance tends to take significantly longer than visual and audiovisual guidance, and leads to a lower precision and a higher subjective workload. In turn, it leaves the visual channel completely unoccupied. Audiovisual guidance often enables the highest precision, while the workload is similar to visual guidance. It allows users to take their visual focus off the monitor towards the patient and their own hands from time to time. These are very promising results that clearly show the potential of sonification in image-guided surgery.

However, all these studies utilized sonifications that only provide very little navigation information. None of the sonifications provides (1) three dimensions that are (2) orthogonal, (3) continuous,
(4) exhibit a high resolution, (5) two polarities (e.g., left and right, front and back), and an (6) absolute coordinate origin.

For example, \cite{davidjournal,surgicalnavigationau,otologic} do provide three discrete distance cues to assist clinicians in marking a cutting trajectory for cancer resection and in avoiding hitting critical nerves and membranes. But the sonifications tend to have discontinuous dimensions, only one polarity, a low resolution and no absolute coordinate origin.  \cite{surgerycoll} provides a one-dimensional sonification with one polarity, indicating the distance to the nearest critical structure to help clinicians avoid harming sensitive membranes, important nerves or large vessels. \cite{surgdistson} describe multiple one-dimensional sonifications with one polarity that should serve as a distance indicator for many kinds of surgical interventions. \cite{surgrev2} provide two-dimensional sonifications with one polarity each, and no absolute coordinate origin, which is why they added a confirmation earcon whenever the target is reached. One sonification helped find the location on the patient's back (simplified to a two-dimensional surface), and a second sonification helped identify the two incision angles in a pedicle screw placement task. \cite{davidstandalone} implemented a two-dimensional sonification that has to take turns with a reference tone in order to provide an absolute coordinate origin. Taking turns with a reference tone interrupts the navigation. Furthermore, one dimension only provides discrete steps and no continuous dimension. The sonification aims at navigating a clinician towards the incision point of a needle on a (more or less) two-dimensional surface, like the abdomen. \cite{pomaziemerblacksch}  provide two continuous dimensions, both with two polarities and an absolute coordinate origin. Again, the contemplated use case is to guide a clinician towards a pre-planned needle-insertion point on a comparably plane surface, like the abdomen.

Motivated by the results of the presented studies, we implemented and experimentally evaluated our psychoacoustic sonification. This is the first study in which the sonification simultaneously provides continuous navigation information about three dimensions, each with two polarities, a high resolution and an absolute coordinate origin. Some surgical procedures cannot be narrowed down to one- or two-dimensional tasks. For such procedures, a three-dimensional sonification is needed.

The surgical use case is: finding bone drilling points on the skull for a craniotomy. In a craniotomy, a constellation of burr holes is drilled at specific points on the skull. After drilling, the holes are connected via manually pulled craniotomy saw wires, so that a region of the skull can be removed. This procedure is often image-guided to ensure that the opening enables removal of the correct portion of the skull, e.g., right above a cerebral edema, or at the optimal location to reach a tumor from a pre-planned angle. This procedure is a three-dimensional (three degrees of freedom) task, where the three-dimensional location on the skull needs to be found, while the orientation of the drill is not crucial.


\subsection{Aim and Research Question}
In the study at hand, we evaluate  how well a three-dimensional sonification can guide users towards a target in comparison to visual and audiovisual guidance.

The potential danger of adding more guidance information to a sonification is that it may overwhelm the user, preventing him or her from navigating to the desired target. But the potential benefit is that it enables clinicians to carry out procedures that cannot easily be narrowed down to one- or two-dimensional tasks, like navigation through a three-dimensional space. Even though quite different in nature, the results of previous studies serve as a reference to judge whether the three-dimensional sonification is an effective guidance tool.

\section{Method}
\label{method}
We invited $24$ non-clinicians to a phantom experiment (mostly computer science and digital media students and staff, age $= 19 \text{ to } 42$, median $= 28$, $19$ male, $5$ female). All participants reported normal hearing and normal vision. We explained the participants the task, showed and described the setup, the auditory, the visual and the audiovisual guidance modes to them and let them explore freely. This introduction and training phase took about one hour, mostly to learn how to interpret the sonification.

\begin{figure}[ht]
\centerline{\includegraphics[width=83mm]{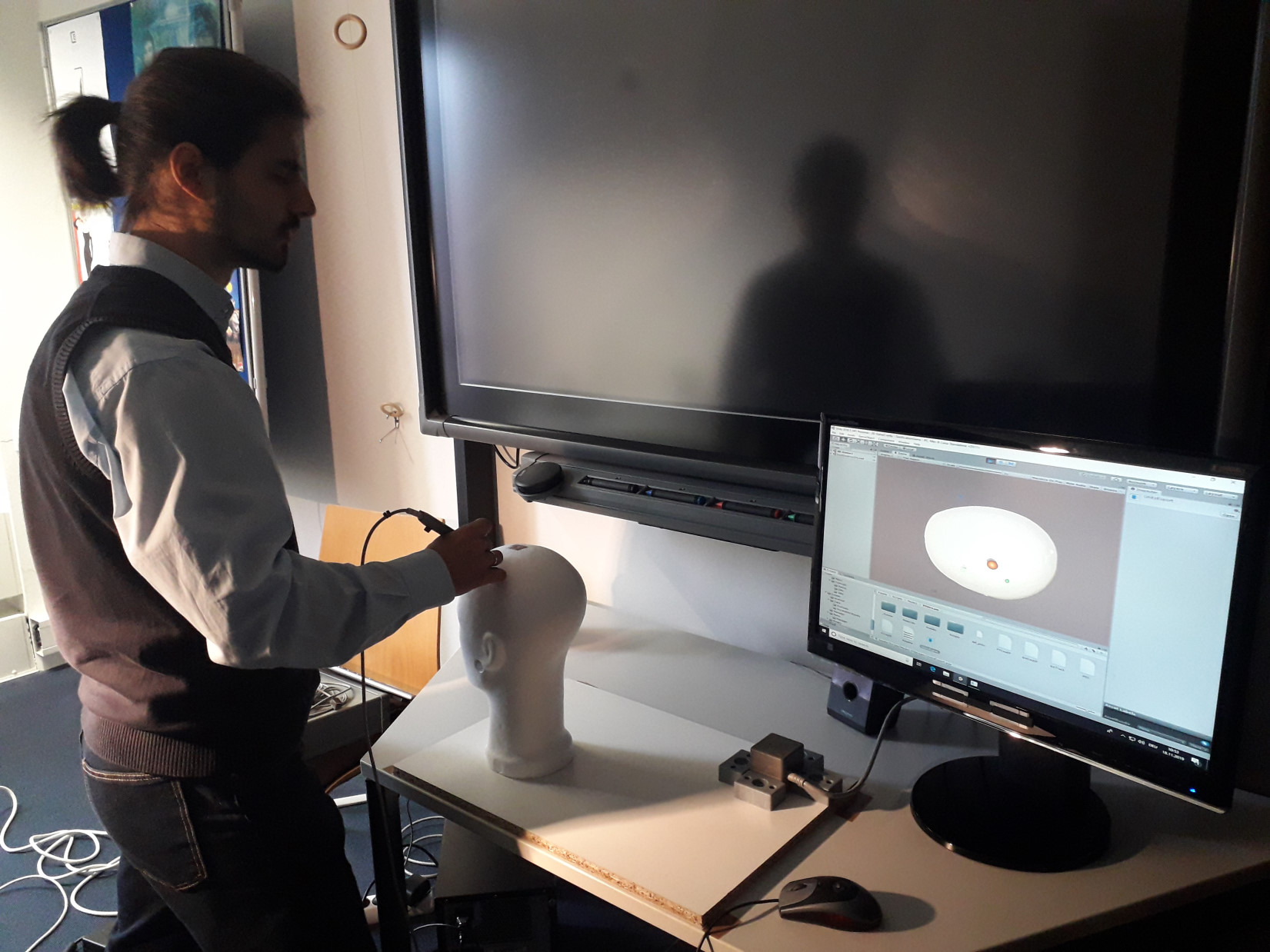}}
	\caption{{\it Experiment setup with a styrofoam skull and an electromagnetic transmitter glued on a board. The position of the stylus tip is tracked. In the auditory guidance mode, the direction between the tip position and the target position is sonified through loudspeakers, and the screen is black. In the visual guidance mode, the tip position and the target is visualized together with a pseudo 3D-model of the skull. In the audiovisual guidance mode, both sonification and visualization are active. Photo taken from \cite{ijis}.}}
	\label{pic:setup}
\end{figure}

\begin{figure}[ht]
\centerline{\includegraphics[width=83mm]{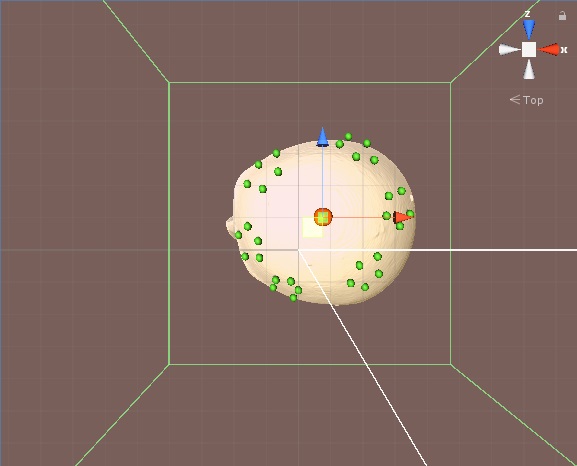}}
	\caption{{\it Distribution of targets along the skull. Six groups of five circularly arranged targets have to be found in pseudo-random order. Only the  target is being sonified and/or visualized. The participants never got to see or hear this distribution. Graphic taken from \cite{icad2023}.}}
	\label{pic:targets}
\end{figure}

The experiment setup can be seen in Fig. \ref{pic:setup}. A styrofoam skull is glued on a board, together with a transducer from the electromegnetic tracking system Polhemus Fastrak. The tracking system is certified for use in operating rooms. We produced a virtual model of the skull using a Siemens MRI scanner. Participants hold a stylus whose tip position is tracked in relation to the skull. Just like a real skull, the phantom hardly provides visual landmarks that could serve as orientation cues. We distributed $6$ times $5$ targets on the surface of the skull as illustrated in Fig. \ref{pic:targets}. The participants' main task was to move the stylus from one target to the next, and click the button on the stylus when they think they have reached it. The path was pseudo-random, but it was the same path for every participant. The participants were not aware of the distribution of targets. They started at one tiny mark on the skull. The guidance method only indicated the position of the target to reach.

Three guidance modes existed: visual guidance (visualization), auditory guidance (sonification) and audiovisual guidance (both). These are described in detail in the subsequent section. In the visual mode (v), the target was indicated as a small, green sphere on the computer model of the skull, while the sonification was muted. In the auditory mode (a), the target location was indicated by the described interactive sonification, while the computer screen was black. In the audiovisual mode (av), both the green sphere and the sonification were active. The participants started with either guidance mode. Every $10$ trials (decade), the guidance method changed. To eliminate the effect of the order of guidance modes, we divided the participants into $6$ different groups, a--v--av, a--av--v, av--a--v, av--v--a, v--a--av, and v--av--a. This way, $8$ participants navigated to each decade of targets using auditory guidance, $8$ using visual guidance, and $8$ using audiovisual guidance. We instructed them to reach the targets as quickly and precisely as possible, which is a common task in surgical experiments \cite{quickprecise}. We tracked the location of the stylus tip during the complete experiment. 

We compare six objective measures between the three navigation modes for each decade of targets:

\begin{enumerate}
    \item Trajectory length
    \item Task duration
    \item Absolute precision and the precision along each of the $3$ dimension
\end{enumerate}


\subsection{Visual Guidance}
\label{visualization}
The visualization could be seen on the photo in Fig. \ref{pic:setup} and in the screenshot, Fig. \ref{pic:targets}. The computer screen shows a two-dimensional projection of the three-dimensional skull model as measured with magnetic resonance imaging. The screen shows a top view of the pseudo three-dimensional skull, comparable to the transverse plane. This was the only viewpoint from which no target was occluded by any part of the skull. Targets were represented by green spheres. The tip of the stylus was represented by a slightly larger, semi-transparent, orange sphere. When moving the stylus to the left or right, the orange sphere would move to the left or right, too. When moving the stylus to the front or back, the orange sphere would move up or down. This is due to the virtual perspective. When moving the stylus to the front or back, the orange sphere may slightly move along the left/right and up/down dimension, and change its size. Participants know that they have reached the target, when the green sphere lies completely within the orange sphere. During the visual guidance, the sonification was automatically muted.

\subsection{Auditory Guidance}
\begin{figure}[ht]
\centerline{\includegraphics[width=83mm]{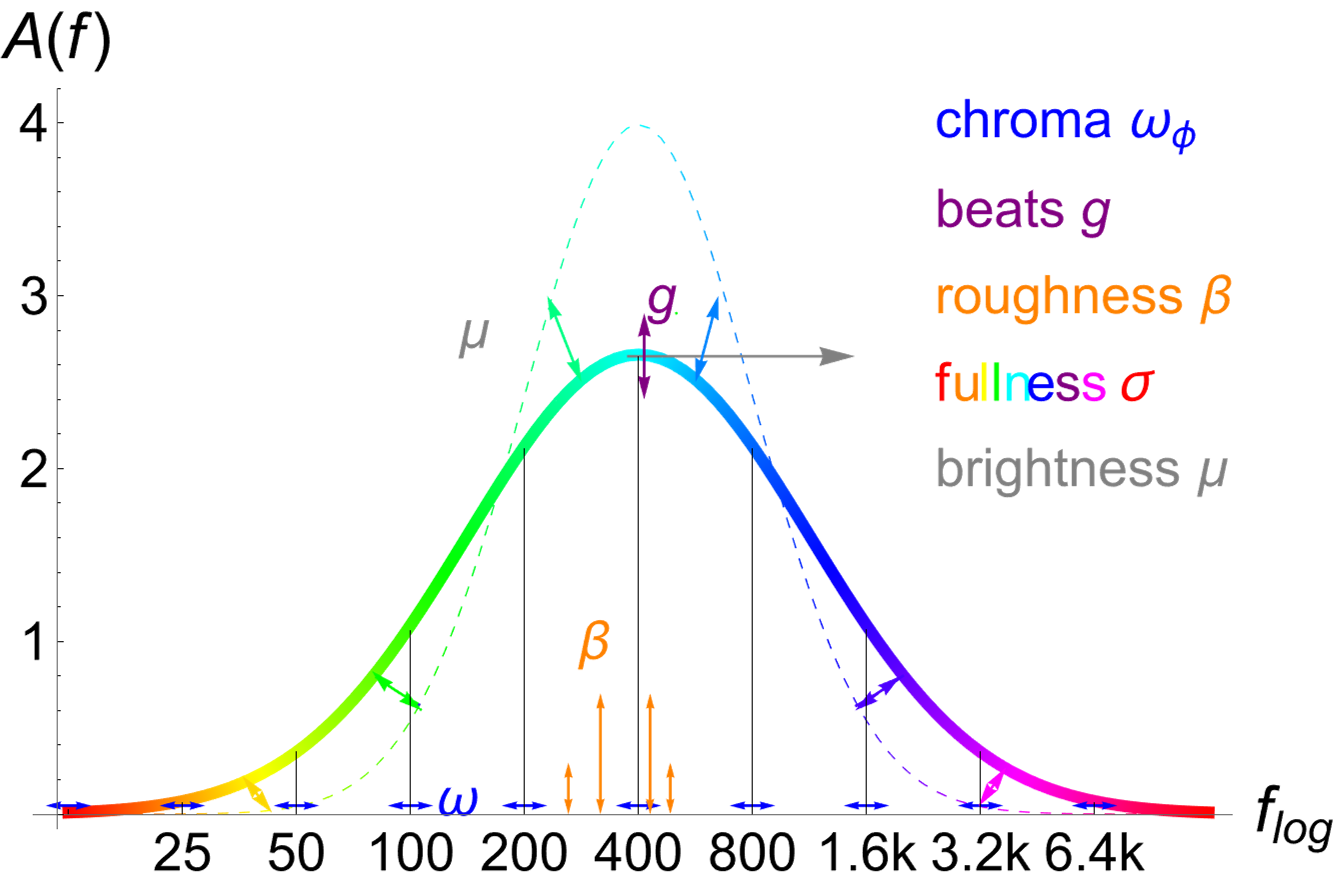}}
	\caption{{\it Magnitude spectrum of the sonification. Five sound characteristics represent the six orthogonal directions in three-dimensional space. The arrows indicate how the spectrum changes when the target lies in either of the $6$ directions. The legend indicates how these changes affect the auditory perception. Graphic modified from \cite{icad2019}.}}
	\label{pic:signi}
\end{figure}

\begin{figure}[ht]
\centerline{\includegraphics[width=83mm]{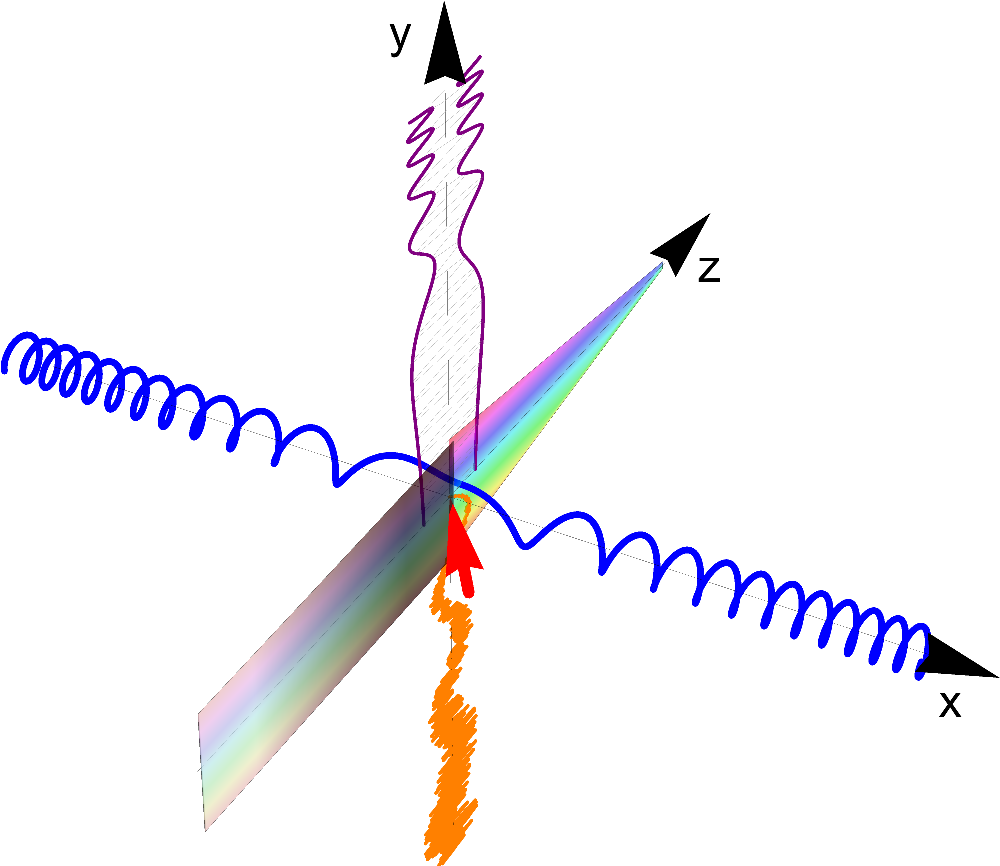}}
	\caption{{\it Sound impression of the sonification depending on the location of the target. Chroma changes indicate that the target lies to the left or right. Loudness fluctuation indicates that the target lies above, roughness indicates that the target lies below. Brightness indicates that the target lies to the rear, a narrow bandwidth indicates that the target lies to the front. Graphic taken from \cite{gfm2020}.}}
	\label{pic:soni}
\end{figure}

A three-dimensional, interactive sonification serves as auditory guidance tool. The sonification is based on our psychoacoustic sonification as introduced for two dimensions in \cite{ziemerschicad} and implemented in the Tiltification spirit level app \cite{tilt}. In \cite{icad2019}, we described how to expand it to a three-dimensional sonification, and in \cite{arxiv} we modified it and provided theoretical and experimental evidence that the three sonification dimensions are in fact continuous, linear, orthogonal, and provide a clear coordinate origin. The sonification is in mono to ensure compatibility with common devices in the operating room, as well as to ensure that clinicians can use it without the need to wear headphones, or the restriction to stay still on the sweet spot of a stereo triangle. The sound is based on additive synthesis, frequency modulation synthesis, amplitude modulation, and subtractive synthesis. In its core, the sonification is a Shepard-Tone \cite{shepardtone1} as illustrated in Fig. \ref{pic:signi}. Here, the vertical black bars indicate the carrier frequencies, the colorful envelope represents the frequency-dependent amplitude. Arrows indicate how frequencies and the envelope can change. These changes affect the perception of chroma, loudness fluctuation or beats, brightness, roughness, and fullness. Each of these characteristics stands for another spatial direction. The intensity of the characteristic, or the frequency at which it varies, indicates how far a target lies along the respective direction. The easiest way to describe the sonification design and the resulting sound is by referring to Fig. \ref{pic:soni}, which represents the perceptual result of the signal changes.

The current position is represented by a cursor in the center of the coordinate system. The left/right dimension is represented by the blue spring, whose coils get denser with increasing distance. When the target lies to the right, all frequencies rise. When the target lies just slightly to the right, the frequencies rise slowly. The further the target lies to the right, the quicker the frequencies rise. When rising slowly, this sounds like an ever-rising pitch. When rising faster, you can hear that something cyclic is happening. This is referred to as a clockwise movement of chroma. Accordingly, when the target lies to the left, all frequencies decrease. When the target lies just slightly to the left, the frequencies decrease slowly. The further the target lies to the left, the quicker the frequencies fall. When decreasing slowly, this sounds like an ever-decreasing pitch. When decreasing faster, you can hear that something cyclic is happening. This is the counter-clockwise movement of chroma. Only when the target lies neither to the left nor to the right, the pitch and chroma remain steady. The exact pitch is not important. Any steady pitch indicates that the target has been approached along the left/right dimension. In Fig. \ref{pic:signi}, a clockwise-motion of chroma looks like a continuous motion of the vertical black bars towards the right, indicated by the blue double arrows ($\omega$). Their length is restricted by the colorful envelope. Whenever a bar leaves the plot on the right, it will be re-introduced on the left-hand side of the plot. The further a target lies to the right, the faster the bars move.

The up/down dimension is divided in two. When the target lies above, the gain of the Shepard-Tone is modulated sinusoidally. This sounds like a loudness fluctuation, also referred to as beats. The further the target lies above, the faster the beating. In Fig. \ref{pic:signi}, this is indicated by the purple double-arrow ($g$), which scales the height of the envelope by a factor that sinusoidally fluctuates between $0.5$ and $1$. The distance to the target is proportional to the frequency of that fluctuation. When the target lies below, all frequencies are modulated with an $80$ Hz modulation frequency. This frequency modulation produces sidebands that give the Shepard-Tone a rough timbre. The further the target lies below, the higher the modulation depth, i.e., the rougher the sound. In Fig. \ref{pic:signi}, this is indicated by the orange arrows ($\beta$) that appear between the vertical bars. In the sonification, these sidebands occur around all carrier frequencies (not only around one frequency, as in the figure). The further the target lies below, the more sidebands occur, and the higher their amplitudes become, increasing the roughness. Only at the target height, the Shepard-Tone exhibits a steady loudness and no roughness. In addition, a click is being triggered every time the target height is surpassed.

The front/back dimension is also divided in two. When the target lies to the front, the bandwidth of the signal is reduced. One can say that this decreases the fullness of the Shepard-Tone, i.e., the sound becomes thinner. The further the target lies to the front, the thinner the respective sound gets. In Figure \ref{pic:signi}, this is indicated by the dashed envelope and the colored double arrows between them. When the target lies far to the front, the Gaussian bell shape becomes much steeper. At the same time, the peak becomes higher. This is necessary, because reducing the sounds' bandwidth would reduce its loudness, which could be confused with beats. Increasing the peak amplitude counter-balances this effect. When the target lies to the back, the amplitude-envelope of the Shepard-Tone is shifted towards higher frequencies, making the sound brighter. The further the target lies to the rear, the brighter the sound gets. In Fig. \ref{pic:signi}, this is indicated by the gray arrow. Only at the target depth, the sound is both full and dull. In addition, a major chord is triggered every time the target depth is surpassed.

Altogether, you know that you have reached the target when the chroma and the loudness are steady and the Shepard-Tone is neither rough nor thin nor very bright. In addition, pink noise is triggered as soon as the target lies nearer than $3$ cm away. During the auditory guidance, the computer screen automatically turned black.

The three-dimensional sonification used in this experiment can be explored in the CURAT sonification game \cite{curat}\footnote{Available for \href{https://curat.informatik.uni-bremen.de/en/download.html}{Android, Windows, Mac and Linux}.}. A demo video of the sonification can be found on \url{https://youtu.be/CbLoQ8LECGw}. The formulas of the digital signal processing are given in \cite{ziemerschicad} and \cite{icad2019}, its implementation in Pure Data is provided in the open source code of Sonic Tilt \cite{sonictilt}.


\subsection{Audiovisual Guidance}
The audiovisual guidance mode was a combination of the visual and the auditory guidance. 

\section{Results}
\label{results}
A visual inspection of the calculated measures showed that the data were not normally distributed but exhibited outliers that led to a large variance, as indicated already in \cite{icad2023}. Therefore, outliers were removed for the data analysis.

We considered data points above $\mathrm{Q3} + 1.5 \text{IQR}$ and below $Q1 - 1.5 IQR$ as outliers and excluded them from further analyses. Here, Q1 is the lower quartile, i.e., the $25$th percentile, Q3 is the upper quartile, i.e., the $75$th percentile, and IQR is the interquartile range, defined as $\text{IQR} = \mathrm{Q3} - \mathrm{Q1}$. For each measure, between $0$ and $4$ values have been excluded this way. Results of the filtered data are presented in the following.

\subsection{Results of Filtered Data}
For each decade, MANOVA with a significance level of $0.05$ and a $95$\% confidence interval is carried out to identify whether the performance difference are significant.
Tables \ref{tab:c1} to \ref{tab:c3} show the mean values $\pm$ standard deviation of all measures for targets $1$ to $10$ (decade $1$), $11$ to $20$ (decade $2$) and $21$ to $30$ (decade $3$). The best performance is highlighted in bold font. The superscripts indicate which guidance modes achieved a significantly better result. With the cleaned data, all MANOVAs revealed significant differences between the guidance modes.

\begin{table}[!ht]
    \centering
    \begin{tabular}{|l|l|l|l|}
    \hline
          & a & v & av \\ \hline
        time &  $271\pm86$ s$^\text{v, av}$ & $\mathbf{52}\pm9$ s & $84\pm48$ s\\ \hline
        length & $176.5\pm46.4$ cm $^\text{v, av}$ & $\mathbf{58}\pm9.9$ cm & $84.7\pm46.7$ cm\\ \hline
        prec & $1.43\pm0.23$ cm$^\text{v, av}$& $0.97\pm0.23$ cm$^\text{av}$& $\mathbf{0.62}\pm0.17$ cm\\ \hline
        $\text{prec}_x$ & $0.78\pm0.20$ cm$^\text{v,av}$ & $\mathbf{0.34}\pm0.05$ cm & $0.39\pm0.08$ cm \\ \hline
        $\text{prec}_y$ & $0.86\pm0.24$ cm$^\text{av}$ & $0.85\pm0.23$ cm$^\text{av}$ & $\mathbf{0.31}\pm0.09$ cm\\ \hline
        $\text{prec}_z$ & $0.80\pm0.13$ cm$^\text{v, av}$ & $\mathbf{0.29}\pm0.17$ cm& $0.33\pm0.22$ cm\\ \hline
    \end{tabular}
    \caption{Mean values of cumulated time and trajectory length, average precision and precision along the $x$-, $y-$, and $z$-axis for targets $1$ to $10$ (decade $1$) after elimination of outliers. Boldface highlights the best guidance method in terms of each measure. The superscripts indicate which guidance mode(s) yielded a significantly better mean value.}
        \label{tab:c1}
\end{table}

The cleaned decade $1$ showed a significant difference ($F(12,22)=3.26$, $p<0.01$, Wilk's $\Lambda=0.031$, partial $\eta^2=0.824$) between the guidance modes. For the test of between-subject effects, we carry out Bonferroni's alpha correction, i.e., the $p<0.01$ significance level is reduced to $p<0.0017$ and the $p<0.05$ significance level to $p<0.008$. The test of between-subject effects revealed that the guidance mode had a significant effect, not only on the time needed ($F(2,16)=26.54$, $p<0.0017$ after Bonferroni correction, partial $\eta^2=0.77$), but also on the trajectory length ($F(2,16)=15.369$, $p<0.0017$, partial $\eta^2=0.66$), the precision along the $x$-dimension ($F(2,16)=23.51$, $p<0.0017$, partial $\eta^2=0.75$), the precision along the $y$-dimension ($F(2,16)=17.72$, $p<0.0017$, partial $\eta^2=0.69$), the precision along the $z$-dimension ($F(2,16)=15.43$, $p<0.0017$, partial $\eta^2=0.66$), and the overall precision ($F(2,16)=24.54$, $p<0.0017$, partial $\eta^2=0.754$). Tukey post-hoc analysis revealed that the time difference is significant between auditory and audiovisual guidance ($p<0.01$) as well as between auditory and visual guidance ($p<0.01$), but not the difference between audiovisual and visual guidance ($p=0.558$). The length difference is significant between auditory and audiovisual guidance ($p<0.01$) as well as between auditory and visual guidance ($p<0.01$), but not the difference between audiovisual and visual guidance ($p=0.453$). The difference of precision along the $x$-dimension is significant between auditory and audiovisual guidance ($p<0.01$), auditory and visual guidance ($p<0.01$) as between audiovisual and  visual guidance ($p<0.01$). The difference of precision along the $y$-dimension is significant between auditory and audiovisual guidance ($p<0.01$) and between visual and audiovisual guidance ($p<0.01)$, but not between auditory and visual guidance ($p=0.987$). The difference of precision along the $z$-dimension is significant between auditory and visual guidance ($p<0.01$) and between auditory and audiovisual guidance ($p<0.01$), but not between visual and audiovisual guidance ($p=0.893$). The difference of overall precision is  significant between all guidance methods ($p<0.01$).

\begin{table}[!ht]
    \centering
    \begin{tabular}{|l|l|l|l|}
    \hline
          & a & v & av \\ \hline
        time &  $160\pm20$ s$^{\mathrm{v}}$ & $\mathbf{51}\pm16$ s & $108\pm81$ s\\ \hline
        length & $110.2\pm11.6$ cm$^{\mathrm{v}}$ & $\mathbf{64.1}\pm16.8$ cm & $91.1\pm22.8$ cm$^{\mathrm{v}}$\\ \hline
        prec & $\mathbf{0.35}\pm.03$ cm& $0.37\pm0.09$ cm& $0.65\pm0.98$ cm\\ \hline
        $\text{prec}_x$ & $1.00\pm0.12$ cm & $\mathbf{0.49}\pm0.28$ cm & $0.87\pm0.58$ cm \\ \hline
        $\text{prec}_y$ & $1.10\pm0.23$ cm$^{\mathrm{av}}$ & $1.04\pm0.26$ cm$^{\mathrm{av}}$ & $\mathbf{0.59}\pm0.43$ cm\\ \hline
        $\text{prec}_z$ & $1.09\pm0.04$ cm & $\mathbf{0.46}\pm0.11$ cm& $1.01\pm0.75$ cm\\ \hline
    \end{tabular}
    \caption{Mean values of cumulated time and trajectory length, average precision and precision along the $x$-, $y$-, and $z$-axis for targets $11$ to $20$ (decade $2$) after elimination of outliers. Boldface highlights the best guidance method in terms of each measure. The superscript indicates which guidance mod(s) yielded a significantly better mean value.}
        \label{tab:c2}
\end{table}

Decade $2$ revealed a significant difference between the guidance modes ($F(12,24)=6.59$, $p<0.01$, Wilk's $\Lambda=0.054$, partial $\eta^2=0.767$). The test of between-subject effects revealed that the guidance mode had a significant effect on the time needed ($F(2,17)=7.65$, $p<0.008$, partial $\eta^2=0.47$) and the trajectory length ($F(2,17)=10.883$, $p<0.0017$, partial $\eta^2=0.56$). The difference in precision along the $y$-axis is close to the desired significance level ($F(2:17)=5.16$, $p=0.018$, partial $\eta^2=0.38$). Tukey post-hoc analysis revealed that the time difference is significant between auditory and visual guidance ($p<0.01$), but not between auditory and audiovisual ($p=0.176$) nor between visual and audiovisual guidance ($p=0.118$). The length difference is significant between auditory and visual guidance ($p<0.01$), but neither between auditory and audiovisual guidance ($p=0.157$), nor between visual and audiovisual guidance ($p=0.453$). The difference in precision along the $y$-axis is significant between auditory and audiovisual guidance ($p<0.05$) and between visual and audiovisual guidance ($p<0.05$), but not between auditory and visual guidance ($p=0.95$).

\begin{table}[!ht]
    \centering
    \begin{tabular}{|l|l|l|l|}
    \hline
          & a & v & av \\ \hline
        time &  $199\pm22$ s$^\text{v, av}$ & $\mathbf{44}\pm9$ s & $63\pm13$ s\\ \hline
        length & $127.2\pm27.6$ cm$^\text{v, av}$ & $\mathbf{54.4}\pm8.2$ cm & $78.8\pm20.1$ cm\\ \hline
        prec & $0.46\pm.09$ cm& $\mathbf{0.36}\pm0.12$ cm& $0.42\pm0.23$ cm\\ \hline
        $\text{prec}_x$ & $1.12\pm0.78$ cm & $0.69\pm0.56$ cm & $\mathbf{0.63}\pm0.65$ cm \\ \hline
        $\text{prec}_y$ & $1.06\pm0.83$ cm$^{\mathrm{av}}$ & $1.02\pm0.48$ cm$^{\mathrm{av}}$ & $\mathbf{0.62}\pm0.48$ cm\\ \hline
        $\text{prec}_z$ & $2.05\pm2.89$ cm$^{\mathrm{av}}$ & $0.80\pm0.71$ cm& $\mathbf{0.61}\pm0.61$ cm\\ \hline
    \end{tabular}
    \caption{Mean values of cumulated time and trajectory length, average precision and precision along the $x$-, $y$-, and $z$-axis for targets $21$ to $30$ (decade $3$) after elimination of outliers. Boldface highlights the best guidance method in terms of each measure. The superscript indicates which guidance mode(s) yielded a significantly better mean value.}
        \label{tab:c3}
\end{table}

Decade $3$ showed a significant difference ($F(12,18)=15.31$, $p<0.01$, Wilk's $\Lambda=0.008$, partial $\eta^2=0.91$) between the guidance modes. The test of between-subject effects revealed that the guidance mode had a significant effect on the time needed ($F(2,14)=178,74$, $p<0.0018$, partial $\eta^2=0.96$), the trajectory length ($F(2,14)=27.28$, $p<0.0018$, partial $\eta^2=0.80$), and the precision along the $y$-axis ($F(2,14)=11.13$, $p<0.0018$, partial $\eta^2=0.61$). Tukey post-hoc analysis revealed that the time difference is significant between auditory and audiovisual guidance ($p<0.01$) as well as between auditory and visual guidance ($p<0.01$), but not the difference between audiovisual and visual guidance ($p=0.055$). The length difference is significant between auditory and audiovisual guidance ($p<0.01$) as well as between auditory and visual guidance ($p<0.01$), but not the difference between audiovisual and visual guidance ($p=0.081$).



\subsection{Summary of Results}
Auditory guidance was significantly slower than visual guidance in all three decades, and trajectories were significantly longer. In one decade, auditory guidance was significantly less precise than visual guidance. In one decade, it was even more precise, but not significantly.

Audiovisual guidance was significantly more precise along the y-axis than visual guidance in all decades, and not significantly slower. In one decade, the precision was significantly higher than in the other guidance modes. In one decade, the trajectory was significantly longer compared to  visual guidance.

All significant findings were supported by a large effect size (partial $\eta^2\geq 0.38$).

\section{Discussion}
\label{discussion}
From the analysis of the results, three aspects turned out to be important to reflect on. These aspects are discussed in the following three sections.

\subsection{Individual performance vs. Guidance Method}
The large variance between individuals concerning the six measures, and the qualitatively observes performance differences discussed in \cite{icad2023}, indicate that for naive users, the individual skills are much more important than the guidance mode. Almost no significant differences between guidance modes could be observed.

However, it is likely that the individual performances of clinicians would vary less, at least for the visual guidance, as they have expertise in the presented task. We take this into consideration by eliminating outliers. Arguably, results of the cleaned data, as discussed in the following sections, transfer better to clinicians.

\subsection{Effectiveness of Auditory Guidance}
\begin{figure}[ht]
\centerline{\includegraphics[width=83mm]{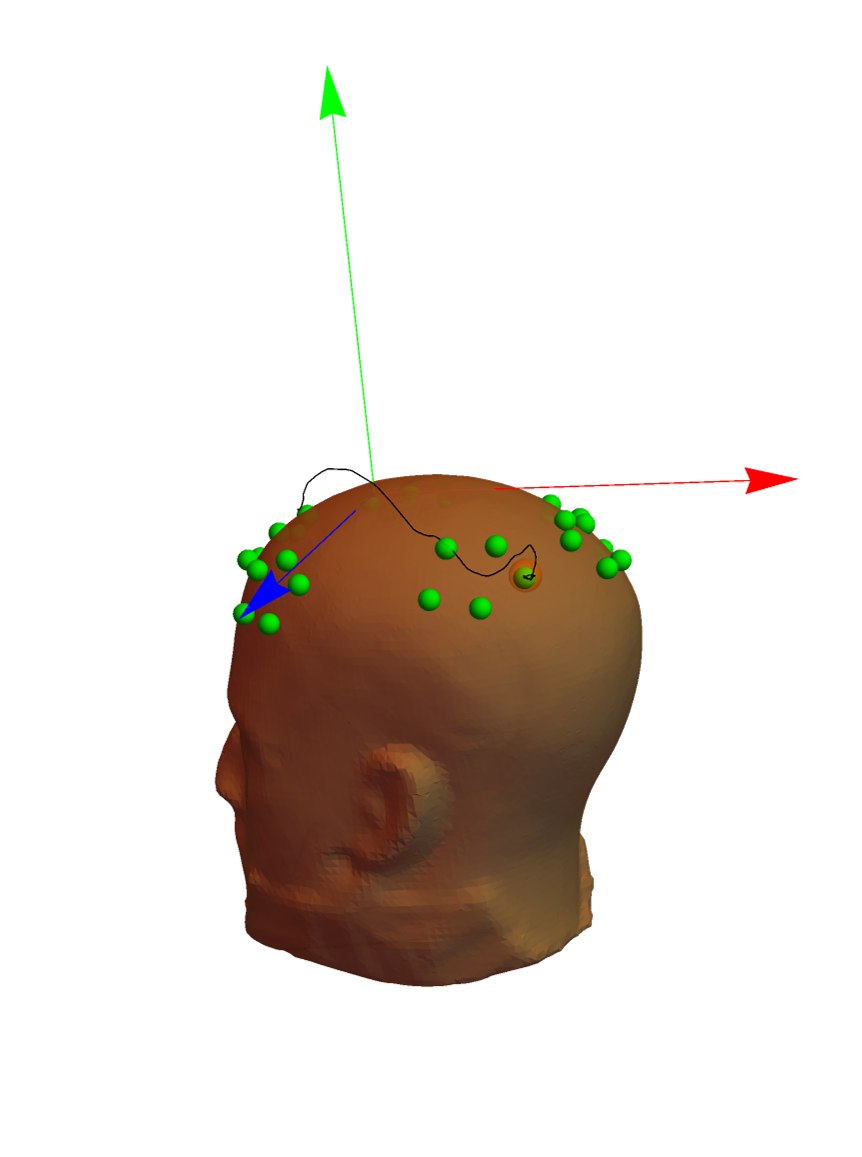}}
	\caption{{\it Exemplary trajectory in the audiovisual guidance mode. Near the target, micro-adjustments are carried out, especially along the $y$-axis. In the experiment, the viewpoint lies above the skull, so the $y$-axis is hardly visible, but distinctly audible during the experiment.}}
	\label{pic:avtarget}
\end{figure}

All participants found all targets, regardless of the guidance mode. Participants tended to perform significantly worse with auditory guidance compared to audiovisual guidance, and, less often, compared to visual guidance. This is true for the time it takes to find the targets, for the length of the trajectories, and for the overall precision. One exception is the precision along the $y$-axis. This is the axis which is hardly recognizable on the screen, due to the virtual viewing angle. Here, audiovisual guidance leads to a significantly higher precision than pure auditory or pure visual guidance. A close look at the trajectories of the audiovisual guidance mode reveals why. As can be seen in Fig. \ref{pic:avtarget}, many participants carried out micro-corrections near the target. Obviously, participants realized that their current position looked right but did not sound right. So they adjusted their position.

Note that the inferior performance using auditory guidance can have various reasons. Maybe the sonification is less informative. But maybe the participants just lack auditory experience and education and can improve their performance with more training. One should keep in mind that users have a life-long experience interpreting visualizations, such as graphical user interfaces on computers, smartphones and TVs, rail network plans and maps and/or pseudo 3D graphics in computer games, but less than one hour of experience using sonification. Longitudinal studies may reveal how much the performance improves after weeks of training.

\subsection{Usefulness of Three-Dimensional Sonification}
In image-guided surgery, some tasks are two-dimensional, such as finding the two desired angles for inserting a needle. Some three-dimensional problems can be simplified as two-dimensional tasks, such as finding a position on a fairly plane surface, like the abdomen. Previous studies have investigated the usefulness of two-dimensional sonifications for such purposes \cite{davidstandalone,surgrev2}.

The fact that the presented results are in line with the results from these studies indicate that adding even more navigation information to the sonification does not degrade its usefulness. Consequently, sonification can even be utilized for three-dimensional guidance tasks without simplifying or subdividing it into two-dimensional subtasks.

One can draw two quite contrary conclusions from the results: 1.) As adding the sonification mostly improved the precision along the y-dimension, it may be beneficial to reduce the sonification to one dimension and add it to the existing visualization. This may improve the navigation precision without the need for much training. After all, the most striking benefit of the presented three-dimensional sonification is the same as the most striking benefit of earlier presented one- or two-dimensional sonifications. 2) One can recognize that the three-dimensional sonification guided all participants through a three-dimensional space with a precision fairly similar to visual guidance. This has not been reported before. It allows speculating that one day sonifying all six degrees of freedom (e.g., three-dimensional location of a needle tip, two-dimensional incision angles and one-dimensional incision depth) will be possible. Until then, longitudinal studies with three-dimensional sonification should reveal with how much training the performance improves by how much.

\section{Conclusion}
\label{conclusion}
In this paper, we described a three-dimensional sonification as an auditory guidance tool for image-guided surgery. In a phantom study, novice users were able to find targets through auditory, visual, and audiovisual guidance. Visual guidance tended to enable the users to find the targets fastest, on the shortest path, and with the highest precision. Adding sonification to the visualization mostly improved the precision along the y-dimension. When replacing visualization completely with sonification, all targets are still found, with a precision that is only significantly lower in one out of three trials.

The result in is line with previous studies that showed the suitability of sonification as a guidance tool in image-guided surgery. The novelty here is that the presented sonification is the first that enabled continuous guidance through a three-dimensional space. The results show that this increase of sonified information is well interpretable by novice users. The results indicate that sonification may become not only a useful supplement, but even a substitute for graphical guidance methods. Yet it is unclear if, and with what amount of training, sonification can become a guidance tool that is as effective as visual guidance. Furthermore, the results clearly show that some people can use sonification more readily than others. It is possible that some clinicians may master sonification guided surgery after a short intensive course and some practice, while others cannot. 
Ultimately, longitudinal studies with clinicians are necessary to reveal the true potential of three-dimensional sonification as an assistance tool for image-guided surgery. Such a study should include qualitative assessments, too, like the subjective confidence, stress, and cognitive demand, and should tackle the integration of sonification in the surgical workflow.
\bmhead{Acknowledgments}
Many thanks to my colleagues for fruitful discussions and advice: Holger Schultheis, Ron Kikinis, David Black, Peter Haddawy and Nuttawut Nuchprayoon
\bibliography{sn-bibliography}
\end{document}